# Coexistence of Tri-Hexagonal and Star-of-David Pattern in the Charge Density Wave of the Kagome Superconductor AV$_3$Sb$_5$

Yong Hu[1,*], Xianxin Wu[2,3,*], Brenden R. Ortiz[4], Xinloong Han[5], Nicholas C. Plumb[1], Stephen D. Wilson[4], Andreas P. Schnyder[2], and Ming Shi[1,*]

[1]Swiss Light Source, Paul Scherrer Institute, CH-5232 Villigen PSI, Switzerland
[2]Max-Planck-Institut für Festkörperforschung, Heisenbergstrasse 1, D-70569 Stuttgart, Germany
[3]CAS Key Laboratory of Theoretical Physics, Institute of Theoretical Physics, Chinese Academy of Sciences, Beijing 100190, China
[4]Materials Department, University of California Santa Barbara, Santa Barbara, California 93106, USA
[5]Kavli Institute of Theoretical Sciences, University of Chinese Academy of Sciences, Beijing, 100049, China

*To whom correspondence should be addressed:
Y.H. (yonghphysics@gmail.com); X.W. (xianxin.wu@fkf.mpg.de); M.S. (ming.shi@psi.ch).

**The recently discovered layered kagome metals AV$_3$Sb$_5$(A=K, Rb, Cs) have attracted much attention because of their unique combination of superconductivity, charge density wave (CDW) order, and nontrivial band topology. The CDW order with an in-plane 2x2 reconstruction is found to exhibit exotic properties, such as time-reversal symmetry breaking and rotational symmetry breaking. However, the nature of the CDW, including its dimensionality, structural pattern, and effect on electronic structure, remains elusive despite intense research efforts. Here, using angle-resolved photoemission spectroscopy, we unveil for the first time characteristic double-band splittings and band reconstructions, as well as the band gap resulting from band folding, in the CDW phase. Supported by density functional theory calculations, we unambiguously show that the CDW in AV$_3$Sb$_5$ originates from the intrinsic coexistence of Star-of-David and Tri-Hexagonal distortions. The alternating stacking of these two distortions naturally leads to three-dimensional 2x2x2 or 2x2x4 CDW order. Our results provide crucial insights into the nature and distortion pattern of the CDW order, thereby laying down the basis for a substantiated understanding of the exotic properties in the family of AV$_3$Sb$_5$ kagome metals.**

Transition-metal-based kagome materials, owing to the unique lattice geometry, provide a fascinating platform to study spin liquid states [1-6], flat bands [7], superconductivity [8-11], charge density wave (CDW) order [9-13] and topological physics [14-17]. Recently, the discovery of a new class of kagome metals $AV_3Sb_5$ (A=K, Rb, Cs), exhibiting topologically non-trivial electronic structure, unconventional superconductivity and CDW order, have been attracting tremendous attention [18-52]. Superconductivity emerges with a critical temperature $T_C$~0.9-2.5 $K$ at ambient pressure [19-21]. Subsequent investigations reveal a significant residual thermal conductivity [27], double superconducting domes under pressure [28-30], and an edge supercurrent in Nb/$K_{1-x}V_3Sb_5$ [31], indicating possible unconventional superconductivity. A CDW phase transition occurs at $T_{CDW}$~78-103 $K$ [19,21,32], for which scanning tunneling microscopy (STM) suggested a broken translation symmetry with a 2x2 in-plane reconstructions [32-34] and broken time-reversal symmetry [32,35]. In addition, an enhanced relaxation rate in the CDW phase from recent zero-field $\mu$SR experiments is in line with a broken time-reversal symmetry [35], which could be intimately related to the observed giant anomalous Hall effect [36,37]. Furthermore, CDW order possessing lattice rotational symmetry breaking [38] can be inferred owing to the two-fold symmetrical c-axis resistivity from magneto-resistance measurements [39]. Theoretical studies provide two scenarios about the in-plane CDW order: phonon softening [40,41] and interaction-driven Fermi surface instability [42-44].

In addition to the in-plane reconstruction, X-ray diffraction (XRD) and STM experiments further reveal modulation along the c-direction [32,45], implying a three-dimensional (3D) CDW order. The charge modulation shows a π-phase shift across the single-unit-cell step edge in STM measurements, suggesting a 2x2x2 CDW order [46]. However, recent XRD measurements suggest a 2x2x4 reconstruction [47] and even a transition between these two by varying temperature [48]. Moreover, the CDW structural pattern is still unclear: NMR measurements tend to suggest a Star of David (SoD) distortion, [49] although the Tri-Hexagonal (TrH) distortion is theoretically more stable [40,41]. Despite the experimental evidence for a 3D CDW order, direct spectroscopic evidence about the reconstruction of the vanadium d-orbitals associated with the kagome lattice in $AV_3Sb_5$ is still lacking, which is crucial for revealing the CDW pattern in the system. A detailed understanding of the 3D CDW structural pattern, and in particular the ensuing electronic structure reconstruction, are of crucial importance for understanding the nature and origin of the CDW order and its interplay with superconductivity.

In this work, we used polarization- and temperature-dependent angle-resolved photoemission spectroscopy (ARPES) to reveal the nature of the CDW in the kagome superconductors AV$_3$Sb$_5$ from an electronic perspective. Upon cooling, the energy location of the two near-$E_F$ VHSs is found to suddenly drop by about 90 *meV*, once the temperature falls below $T_{CDW}$. Moreover, universal band-splitting-like features on the kagome bands are revealed, exhibiting prominent band-folding effects in the kagome lattice, which can be well captured by the DFT calculations including superimposed Tri-Hexagonal and Star-of-David CDW patterns. The characteristic splittings from the two patterns can be experimentally extracted, and they are almost quantitatively consistent with calculations, clearly demonstrating intrinsic coexistence of two patterns in the CDW order. These findings not only provide essential insights into the band reconstructions of the CDW order but also unambiguously identify the 3D nature and CDW distortion pattern, which will be crucial in illuminating its mechanism.

The pristine phase of CsV$_3$Sb$_5$ crystalizes in a layered crystal structure with the space group P6/mmm (No. 191) (Fig. 1a). It consists of V$_3$Sb kagome layers bounded above and below by Sb honeycomb layers and Cs hexagonal networks [Fig. 1a(i)]. Below the CDW transition temperature ($T_{CDW}$ = 94 *K*), the kagome layer can exhibit two types of distortions derived from breathing phonon modes [39]: a Star of David [Fig. 1b(i)] and its inverse structure (Tri-Hexagonal) [Fig. 1b(ii)]. The orbital-resolved band structure of pristine CsV$_3$Sb$_5$ from DFT calculations is shown in Fig. 1c, where Sb $p_z$ orbital contributes an electron band around the Brillouin zone (BZ) center and vanadium *d* orbitals dominate the Dirac cone (DC) bands around the K point and fourfold van Hove singularity (VHS) bands around the M point with diverse sublattice features. To demonstrate the effect of CDW order on the electronic structure, we first compare the ARPES spectra of CsV$_3$Sb$_5$ in the normal state ($T \gg T_{CDW}$) and CDW phase ($T \ll T_{CDW}$). Figure 1d plots the constant energy contours (CECs) and their evolution with binding energy ($E_B$). With increasing $E_B$, the Sb contributed circular-shaped pocket near the zone center ($\Gamma$) and triangle-shaped Fermi surface (FS) derived from the vanadium kagome lattice around the $\bar{K}$ points shrink [(Fig. 1d(i-iii))], reflecting their electron-like nature. In the CDW states ($T \ll T_{CDW}$), one prominent change is that the electron pocket around $\bar{\Gamma}$ is doubled [Figs. 1b(i-iii)], which has been suggested to originate from the $k_z$ broadening of the Sb $p_z$ orbital [50], quantum well states on the surface [51], or the out-of-plane band folding [52]. Moreover, the triangle-shaped CECs associated with the vanadium *d* orbital around $\bar{K}$ enlarge slightly in the CDW phase [Figs. 1e(i-iii)], compared with the corresponding CECs at the same $E_B$ in the normal state [Fig. 1d(i-iii)]. To further identify the effect of CDW order, we compare the band dispersions along the $\bar{\Gamma}$ - $\bar{K}$ - $\bar{M}$ - $\bar{\Gamma}$ direction above and below the $T_{CDW}$. ARPES spectra reveal a downward shift of the bands around $\bar{M}$ point in the CDW phase [Figs. 1f(i) and 1g(i)], which is more evident with linear vertically (*LV*) polarized light [see the arrows in Figs. 1f(ii) and 1g(ii)]. The Dirac-like crossing point between $\bar{\Gamma}$ and $\bar{K}$ point also shifts down significantly in energy in the CDW state, as indicated by the arrows in Figs. 1f(iii) and 1g(iii) (for details, see Supplementary Fig. S1). Since there are multiple VHSs located around the $\bar{M}$ point (Fig. 1c) [24], carrying large density of states, the CDW-induced band reconstruction on them could be prominent.

To further demonstrate the salient features of the band reconstruction at the VHSs in the CDW phase, we performed systematic temperature- and polarization-dependent measurements on $CsV_3Sb_5$ along the $\bar{\Gamma}$ - $\bar{K}$ - $\bar{M}$ - $\bar{K}$ and $\bar{\Gamma}$ - $\bar{M}$ - $\bar{\Gamma}$ directions (Fig. 2). Upon cooling, the overall evolution of the band structures can be divided into two distinct regions: above and below the $T_{CDW}$. As show in Fig. 2a, above $T_{CDW}$ (*T* = 200 *K*, 110 *K*), the band dispersion remains almost unchanged (Figs. 2a and 2b), while once below $T_{CDW}$ (*T* = 87 *K*, 20 *K*), the pronounced band reconstruction around $\bar{M}$ can be observed [Figs. 2c and 2d, as highlighted by the pink box in Fig. 2d(i)]. To clearly disentangle the bands and keep track of the energy location of the VHSs, we show the polarization dependence of the spectra in Figs. 2e-2l. Cooling from the normal state (*T* >> $T_{CDW}$) to the CDW phase (*T* << $T_{CDW}$), VHS1 band exhibits unusual spectral broadening and energy shift along the $\bar{K}$ - $\bar{M}$ direction in the CDW phase [Figs. 2e-2f(i)], which is further accentuated by the temperature-dependent energy distribution curves (EDCs) taken near the flat feature of VHS1 (Fig. 2m). Under our ARPES geometry, VHS2 bands ($d_{yz}$ orbital) along the $\bar{\Gamma}$ - $\bar{K}$ [Figs. 2e-2h(i)] and $\bar{\Gamma}$ - $\bar{M}$ [Figs. 2e-2h(ii)] directions are detected by linear horizontal (*LH*) and *LV* polarized light, respectively [24], while the VHS4 bands with $d_{xz}$ character along the $\bar{\Gamma}$ - $\bar{K}$ [Figs. 2i-2l(i)] and $\bar{\Gamma}$ - $\bar{M}$ [Figs. 2i-2l(ii)] paths are favored under the *LH* and *LV* polarizations, respectively (for details, see Supplementary Fig. S2). The energy location of VHS2 (resp. VHS4) is determined by the $d_{xz}$ band top (resp. $d_{yz}$-band bottom) along $\bar{\Gamma}$ - $\bar{K}$ and its band bottom (resp. band top) along $\bar{\Gamma}$ - $\bar{M}$ direction. Figure 2n shows the temperature dependence of the energy locations of VHS2 and VHS4, which suddenly drop in energy across the CDW transition by about 90 *meV*. Notably, the dichotomy between VHSs bands with different V *d*-orbital characters (Figs. 2m and 2n) indicates that the band reconstruction in the CDW phase is orbitally-selective (also see Supplementary Fig. S3). The band shifts take place suddenly when the system enters the CDW phase (Fig. 2), suggesting that they are intimately related to the CDW order and the CDW transition is first order.

Besides the energy shift of the VHSs, we further display the characteristic double-band splitting features on the vanadium $d$-orbitals of the CDW phase in detail along high-symmetry paths (Figs. 3a-3f). In Figs. 3a and 3b, we show the temperature dependence of the ARPES spectra collected under circular ($C$) polarization along the $\bar{\Gamma}$ - $\bar{K}$ - $\bar{M}$ direction. Remarkably, compared to the bands measured above $T_{CDW}$ ($T$ = 110 K, 200 K), the spectra taken below $T_{CDW}$ ($T$ = 20 K, 87 K) exhibit diverse band-splitting-like features on the $d$-orbital bands in almost the entire energy-momentum space [indicated by the red dashed line and arrow in Fig. 3b(i)]. Specifically, splittings occur near the Fermi level (BS1), on the lower branch of the Dirac cone band along the $\bar{\Gamma}$ - $\bar{M}$ direction (marked as BS2, B3), around the VHS3 bands (BS4), and even on the bands with an $E_B$ lower than 1.0 $eV$ (BS5, BS6). Consistently, the double-band splittings can also be clearly seen in the bands measured at low temperature (Figs. 3d and 3e), along the $\bar{\Gamma}$ - $\bar{M}$ direction [marked as BS7-BS9 in Fig. 3e(i), for details, see Fig. S3]. By showing the ARPES spectra below and above $T_{CDW}$ in the same plot, Figures 3c and 3f highlight the marked contrast between the bands in the normal state and CDW phase. Besides the double-band splitting features, below the $T_{CDW}$, CDW folding gaps of the $d_{xy}$ band can be identified along the $\bar{\Gamma}$ - $\bar{K}$ direction [indicated by the blue arrow in Fig. 3c(ii)], which is consistent with the observation in KV$_3$Sb$_5$ [25]. Moreover, we have checked the band structure of RbV$_3$Sb$_5$ in the CDW phase, and found a similar band reconstruction (i.e., the double-band splittings and folding gaps, see Supplementary Fig. S4 for details). Therefore, the temperature- and material-dependent ARPES spectra (Figs. 3a-3e and Fig. S4) strongly demonstrate that the observed double-band splitting features are directly derived from the CDW order. However, these features cannot be explained by an in-plane CDW order. As shown Fig. 3g, we show the unfolded band structure of CsV$_3$Sb$_5$ by considering the 2×2 TrH reconstruction [Fig. 1b(ii)]. While the two-dimensional (2D) lattice reconstruction can only qualitatively capture the folding gap along the $\bar{\Gamma}$ - $\bar{K}$ direction [see the blue arrow in Figs. 3c(ii) and 3g], the new emergent bands observed in the CDW phase are completely missing.

Motivated by the suggested 2x2x2 CDW order from recent XRD and STM measurements in AV$_3$Sb$_5$ kagome metals, we further explore the band reconstructions from a 3D CDW order. Actually, the

SoD- or TrH-like distortion in the vanadium kagome layers can induce an $A_{1g}$ out-of-plane distortion of the Sb atoms, which naturally introduces interlayer coupling between kagome layers and generates a genuine 3D CDW distortion. For the 2x2x2 CDW order, consisting of SoD- or TrH-like distortions, there are four possible configurations along the *c* axis: SoD-π, TrH-π, SoD-TrH-π and SoD-TrH, where π denotes an in-plane π-phase shift between adjacent kagome layers (see Fig. S5 for details). It is noted that the six-fold rotation symmetry is broken in the former three configurations and the point group is reduced to $D_{2h}$, while all symmetries of the original lattice are preserved in the last one. The unfolded band dispersions for the SoD-π and TrH-π configurations are displayed in Figs. 4a and 4b, respectively (dispersion along other paths can be found in Supplementary Fig. S6 and S7, and there is no qualitative difference). In the SoD-π configuration, the VHS1, VHS2 and VHS4 bands around $\overline{M}$ point are gapped (indicated by the green arrow in Figs. 4a and 4b). In addition, two folding gaps for the $d_{xy}$ band along the $\overline{\Gamma}$ - $\overline{K}$ direction above and below the band crossing can be clearly identified (marked as the black arrow in Figs. 4a and 4b). The band reconstruction in the TrH-π configuration is similar but with much larger splitting and gap, especially for the VHS1 and VHS2 bands. However, these band reconstructions are similar to those from 2D SoD or TrH distortions except for some noticeable band folding around $\overline{\Gamma}$, and thus cannot explain the doubled-splitting features in experiments. We further display the unfolded band dispersion for the structures with alternating SoD- and TrH-like distortions in Figs. 4c (SoD-TrH-π) and 4d (SoD-TrH). The band dispersion in both cases is relatively similar and the band reconstruction consists of contributions from both SoD- and TrH-like distortions, generating double-splitting features. Interestingly, the splitting of the lower Dirac cone bands along the $\overline{K}$ - $\overline{M}$ can be clearly identified (marked as BS2 in Figs. 4d and 4e). By comparing Fig. 4e and Figs. 4c,4d, we find that the observed band reconstructions (BS1, BS2, BS4, BS5, BS7-BS9) in our ARPES experiments can be well reproduced from these two distortions. As the SoD- and TrH-like distortions exhibit distinct splittings and folding gaps, we can further extract the energy scales of the splitting $\Delta_1$ around $\overline{M}$ point, the CDW folding gap $\Delta_2$ along the $\overline{K}$ - $\overline{M}$ direction, the flat $d_{xy}/d_{x2-y2}$-band splitting $\Delta_3$, and the near-$E_F$-band splitting $\Delta_4$, for CsV$_3$Sb$_5$ in experiments. We summarize the experimental values in Table I, in comparison to theoretical values for four different configurations. It is striking that the experimental values are in very good agreement with the theoretical values of the SoD-TrH-π and SoD-TrH configurations. Intriguingly, the double-band splitting on the $d_{xy}$ band along the $\overline{\Gamma}$ - $\overline{K}$ direction (BS3 and BS6) can be better captured by the SoD-TrH configuration. However, this feature can be also reproduced along other paths in the SoD-TrH-π configuration (for details, see Supplementary Fig. S7), where the six-fold rotational symmetry is broken. Therefore, from our ARPES measurement, both configurations are possible, which is further supported by their close energies in theoretical calculations (less than 1 *meV*/f.u.).

Table I: The experimentally determined energy scales of the double-band splittings, and their comparison with theoretical values.

|  | Exp. | SoD-π | TrH-π | SoD-TrH-π | SoD-TrH |
|---|---|---|---|---|---|
| $\Delta_1^{SoD}$ | 70 *meV* | 105 *meV* |  | 130 *meV* | 90 *meV* |
| $\Delta_1^{TrH}$ | 200 *meV* |  | 220 *meV* | 230 *meV* | 210 *meV* |
| $\Delta_2^{SoD}$ | 120 *meV* | 120 *meV* |  | 210 *meV* | 160 *meV* |
| $\Delta_2^{TrH}$ | 170 *meV* |  | 210 *meV* | 210 *meV* | 220 *meV* |
| $\Delta_3$ | 110 *meV* | 120 *meV* | 270 *meV* | 170 *meV* | 130 *meV* |
| $\Delta_4$ | 90 *meV* | 60 *meV* | 240 *meV* | 100 *meV* | 100 *meV* |

Our ARPES measurements, combined with DFT calculations, for the first time, unambiguously reveal band reconstructions originating from both SoD- and TrH-like distortions. This demonstrates the coexistence of the two distortions in the CDW phase, although the SoD phase is theoretically believed to be metastable. These distortions lead to strong reconstructions of both the VHS bands and the vanadium *d*-orbital bands (especially for $d_{xy}/d_{x2-y2}$ bands). The stacking of the two distortions along the *c* axis can generate a 3D CDW order. Their alternating stacking yields a 2x2x2 CDW order. More complicated stacking, such as $TS\overline{TS}$ and $TTST$, where $\overline{T}$ or $\overline{S}$ denotes an in-plane π shift, can quadruple the unitcell along the *c* axis and lead to a 2x2x4 CDW order (see Supplementary Fig. S8 for the bands structure of $TS\overline{TS}$ stacking), which is suggested by recent XRD measurements [47,48]. In the CDW phase, the experimental transition between 2x2x2 and 2x2x4 reconstructions by varying temperature could be related to the change of distortion stacking along the *c* axis [48]. One important feature of a 3D CDW order is that the six-fold rotational symmetry can be broken for the distortions involving an in-plane π shift between adjacent layers. The SoD-TrH-π configuration can account for the observed two-fold symmetrical *c*-axis resistivity in experiments [39]. In addition, the 3D CDW order can also generate band folding along the *c* axis, providing a qualitatively better account for the double electron bands around the $\overline{\Gamma}$ point (Figs. 1e and 1g; Figs. 3c and 3f). While the DFT calculations based on static lattice distortions go a long way in reproducing our band structure measurements in the CDW phase, they do not account for the time-reversal symmetry breaking seen in other experiments [32,35]. This suggests that electron-phonon coupling and electron-electron interactions may conspire to generate the unconventional CDW order in AV$_3$Sb$_5$ kagome metals, which deserves further investigations, both from the theoretical and the experimental side.

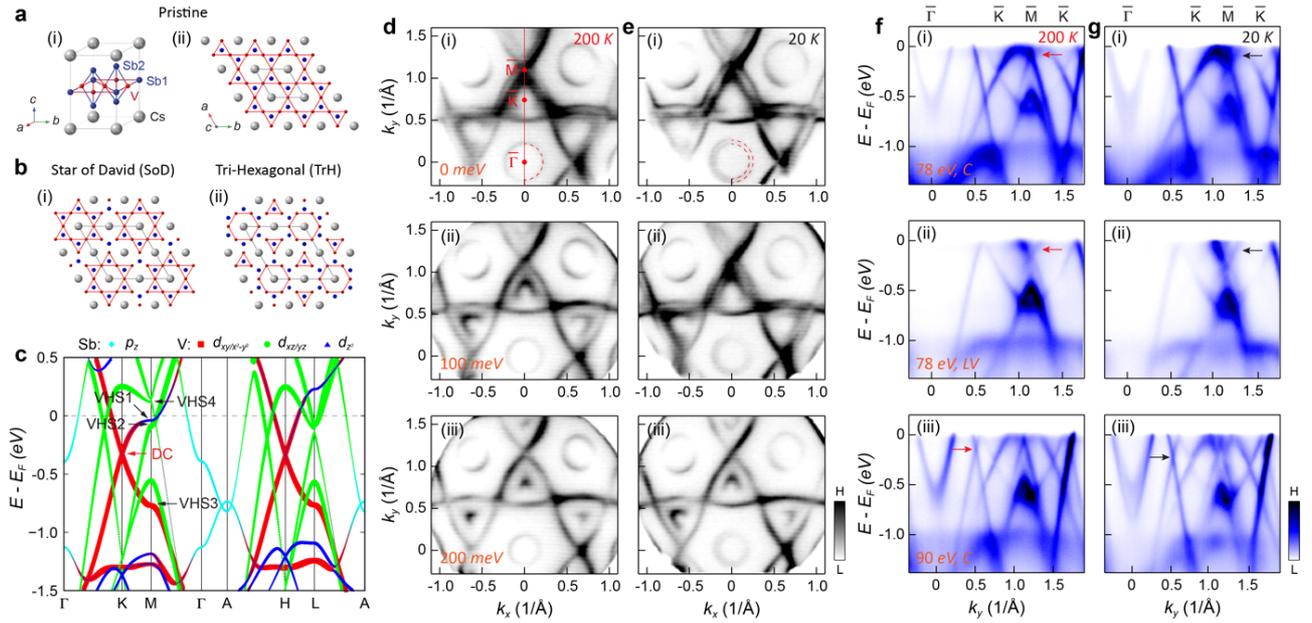

**Fig. 1| Crystal and electronic structures of the kagome superconductors CsV$_3$Sb$_5$ above and below $T_{CDW}$. a** Crystal structure of CsV$_3$Sb$_5$ in the normal state ($T > T_{CDW}$), showing the unit cell in 3D (i), along with the kagome arrangement of vanadium atoms in the *a-b* plane (ii). **b** The crystal structure in the CDW phase ($T < T_{CDW}$) with the candidate Star of David structure (SoD, i) and Tri-Hexagonal structure (TrH, ii). **c** DFT calculations for the orbital character resolved band structure of CsV$_3$Sb$_5$. The orange arrow indicates the Dirac cone (DC) at the K point. **d** Fermi surface (i), constant energy contours at binding energies ($E_B$) of 100 *meV* (ii), and 200 *meV* (iii), at 200 *K*. **e** Same as (d), but taken at 20 *K*. The red dashed line is the guide to the eye for the electron pocket contributed by the Sb $p_z$-orbital. **f** ARPES spectra obtained at 200 *K*, measured with 78 *eV* circular (*C*) polarization (i), 78 *eV* linear vertical (*LV*) polarization (ii), and 90 *eV* C polarization (iii), taken along the $\bar{\Gamma}$ - $\bar{K}$ - $\bar{M}$ - $\bar{K}$ direction as indicted by the red line in [d(i)]. **g** Same as (f), but measured at 20 *K*. The red and black arrows in (f) and (g) highlight the distinct band dispersions above and below the $T_{CDW}$, respectively.

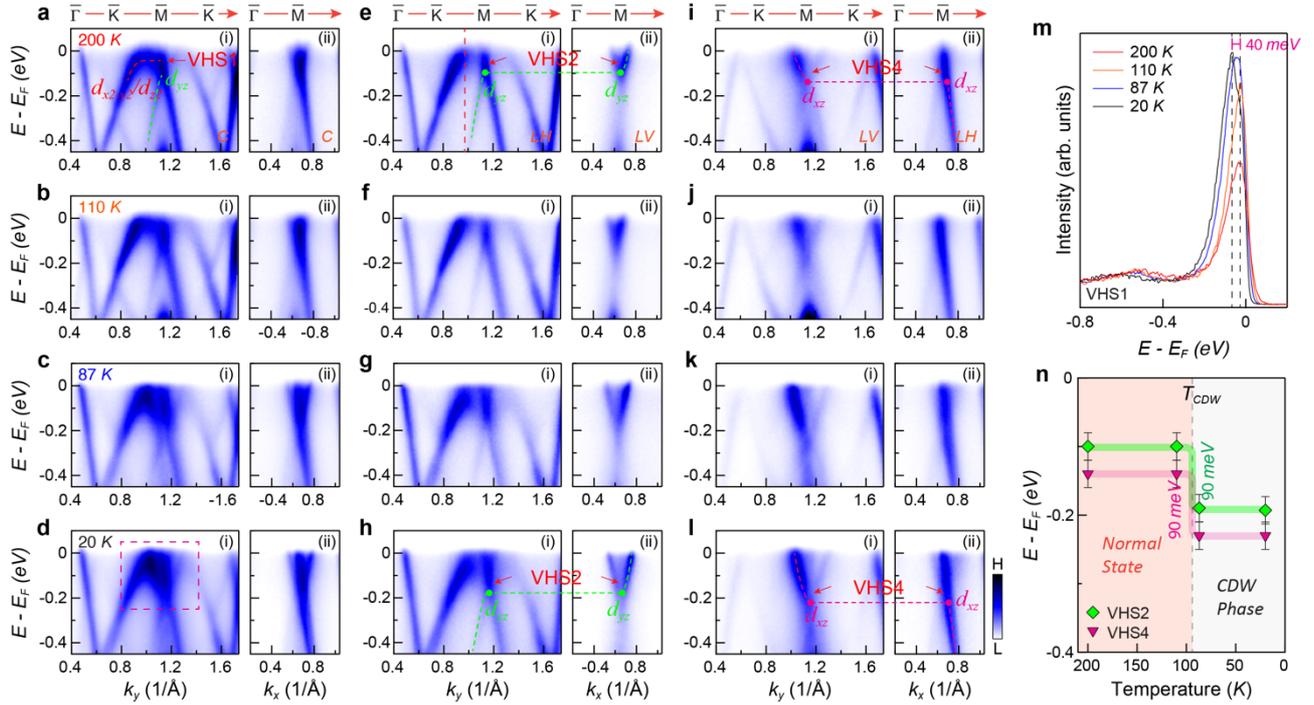

**Fig. 2 | Temperature and polarization dependence of the electronic structure in CsV$_3$Sb$_5$. a–d** ARPES spectra taken along the $\bar{\Gamma}$ - $\bar{K}$ - $\bar{M}$ - $\bar{\Gamma}$ [left panel, (i)] and $\bar{\Gamma}$ - $\bar{M}$ - $\bar{\Gamma}$ [right panel, (ii)] directions, respectively, measured at 200 K (a), 110 K (b), 87 K (c), 20 K (d) with 78 eV C polarized light. **e–h, i–l** Same as (a–d), but obtained with linear horizontal (*LH*) [e-h(i), i-l(ii)] and *LV* [e-h(ii), i-l(i)] polarizations. The dashed curves in (a, e, h, i, l) are guides to the eye for the kagome bands around the $\bar{M}$ point. The red arrow indicates the VHSs. **m** Temperature evolution of the energy distribution curves (EDCs) taken around the flat feature of VHS1. The position of the EDCs is indicated by the red dashed line in e(i)]. **n** Energy location of VHS2 and VHS4 as a function of temperature.

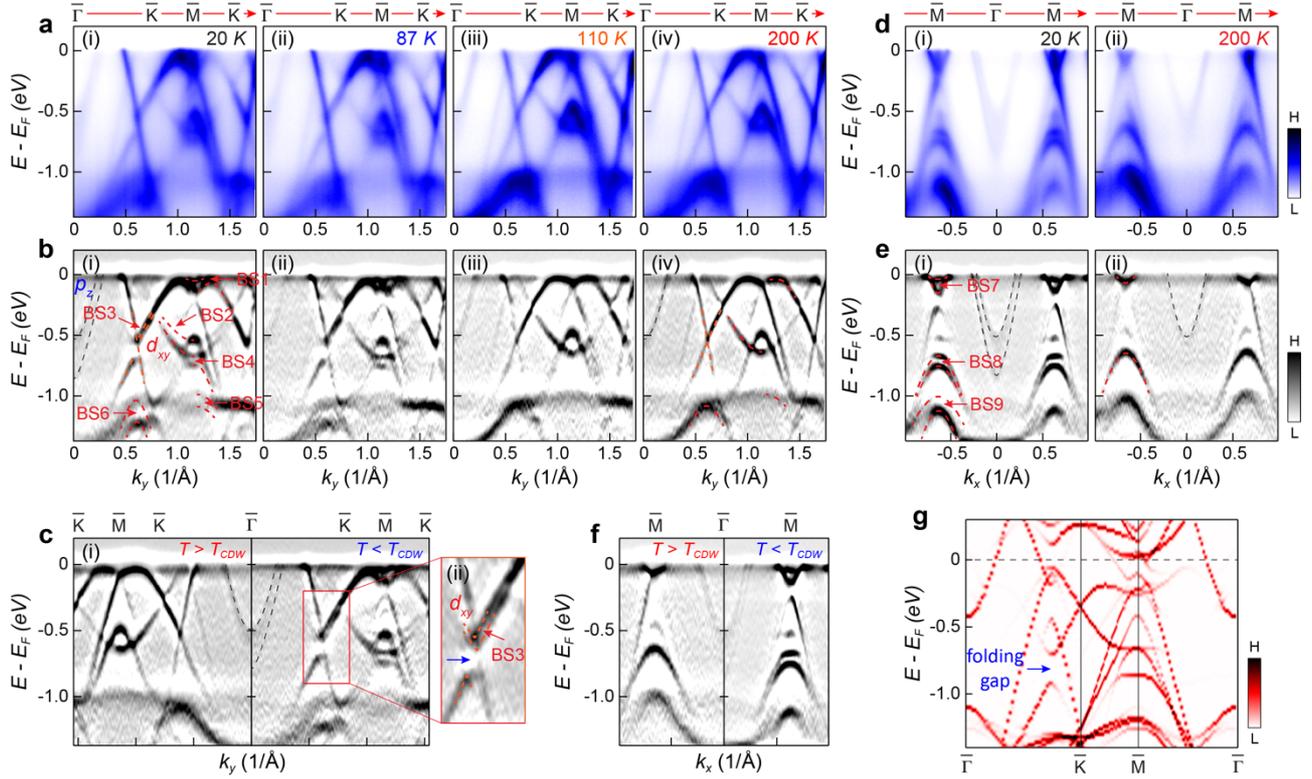

**Fig. 3| Double-band splitting features of the kagome bands below $T_{CDW}$. a,b** ARPES spectra (a) and their curvature plots (b) at different temperatures, taken along the $\bar{\Gamma}$ - $\bar{K}$ - $\bar{M}$ direction. **c** Comparison of the curvature plots taken above $T_{CDW}$ (200 K) and below $T_{CDW}$ (20 K). Orange box in (i) highlights the folding gap on the $d_{xy}$ orbital (ii). **d-f** Same as (a-c), but measured along the $\bar{\Gamma}$ - $\bar{M}$ direction. The green and red dashed lines are guides to the eye for the kagome bands in the normal state and CDW phase, respectively. The red arrows indicate the double-band splittings on kagome bands in the CDW phase, marked as BS1-BS9. **g** DFT calculations for the unfolded band structures of 2x2 TrH phase. The blue arrow in [c(ii)] and (g) indicate the CDW associated folding gap.

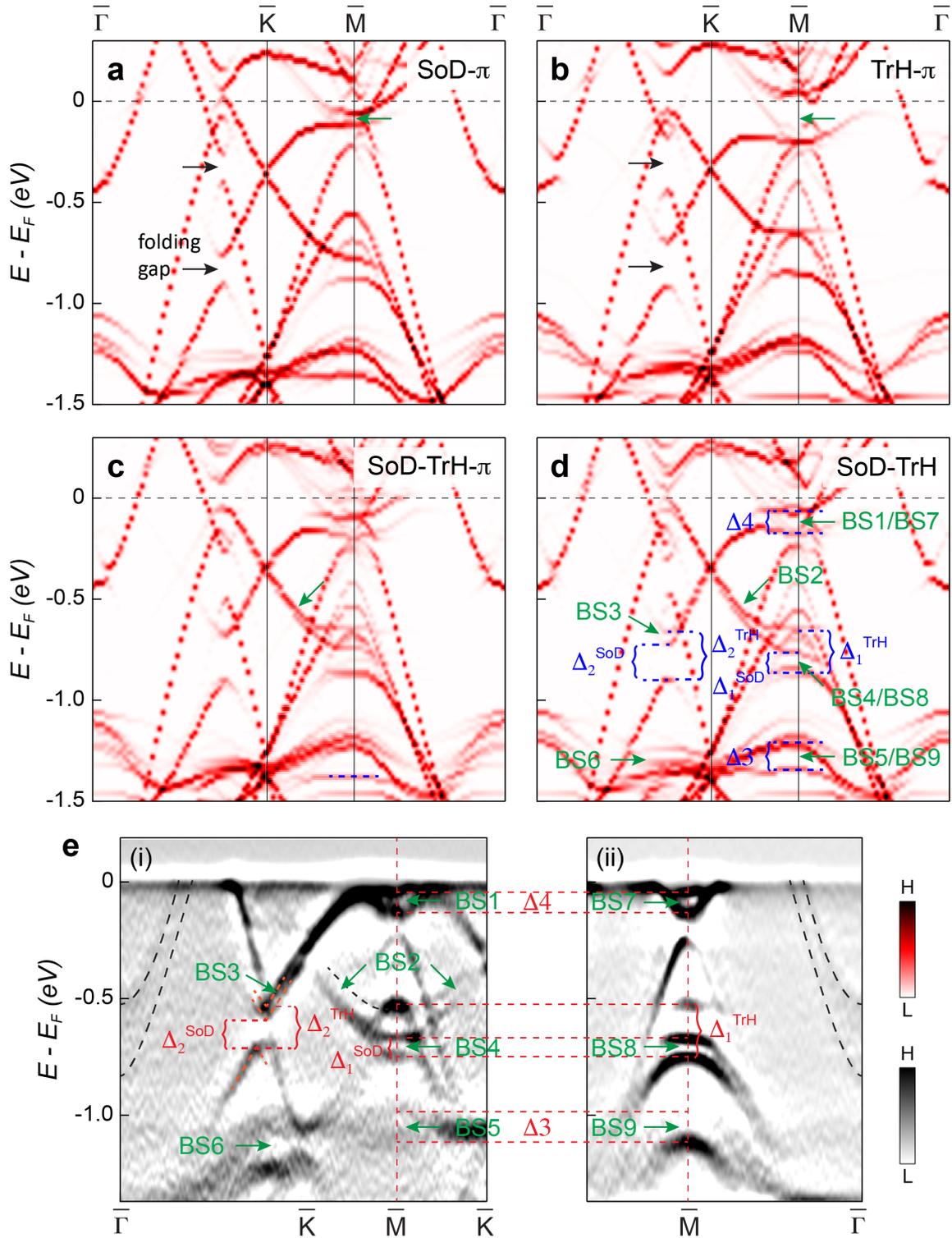

**Fig. 4 | Unfolded band structures of CsV$_3$Sb$_5$ in 2x2x2 CDW order from DFT calculations and their comparison with experiments. a–d** The unfolded band structures of the following lattice configurations: SoD-π (a), TrH-π (b), SoD-TrH-π (c), and SoD-TrH (d). The blue dashed lines define the splitting gaps (($\Delta_1$, $\Delta_3$, $\Delta_4$,) and folding gaps ($\Delta_2$). The superscripts SoD and TrH in the gaps ($\Delta_1$, $\Delta_2$) represent the contributions of the two configurations, respectively. **e** ARPES spectra taken at 20 K along the $\bar{\Gamma}$ - $\bar{K}$ - $\bar{M}$ direction (i) and the $\bar{\Gamma}$ - $\bar{M}$ direction. The green arrow indicates the observed band reconstructions (marked as BS1-BS9) in the CDW phase.